# Atomistic calculations of charged point defects at grain boundaries in $SrTiO_3$

Cong Tao[1,2], Daniel Mutter[1], Daniel F. Urban[1], and Christian Elsässer[1,3]

[1]Fraunhofer IWM, Wöhlerstraße 11, 79108 Freiburg, Germany

[2]Institute of Applied Materials-Computational Materials Science (IAM-CMS), Karlsruhe Institute of Technology, Straße am Forum 7, 76131 Karlsruhe, Germany

[3]Freiburg Materials Research Center (FMF), University of Freiburg, Stefan-Meier-Straße 21, 79104 Freiburg, Germany

## Abstract

Oxygen vacancies have been identified to play an important role in accelerating grain growth in polycrystalline perovskite-oxide ceramics. In order to advance the fundamental understanding of growth mechanisms at the atomic scale, classical atomistic simulations were carried out to investigate the atomistic structures and oxygen vacancy formation energies at grain boundaries in the prototypical perovskite-oxide material $SrTiO_3$. In this work, we focus on two symmetric tilt grain boundaries, namely $\Sigma5$ (310)[001] and $\Sigma5$ (210)[001]. A one-dimensional continuum model is adapted to determine the electrostatic potential induced by charged lattice planes in atomistic structure models containing grain boundaries and point defects. By means of this model, electrostatic artifacts, which are inherent to supercell models with periodic or open boundary conditions, can be taken into account and corrected properly. We report calculated formation energies of oxygen vacancies on all the oxygen sites across boundaries between two misoriented grains, and we analyze and discuss the formation-energy values with respect to local charge densities at the vacant sites.

# 1 Introduction

Both atomic point defects and extended crystallographic defects play significant roles for the physical properties of ceramic materials. For example, various kinds of grain boundaries (GBs) in perovskite-type oxide compounds have been extensively investigated by experiments [1–3] since they are assumed to be responsible for the electrical behavior of the ceramics, as e.g. the dielectric response [4] or thermoelectric resistance [5,6]. Atomistic structures as well as formation energies of symmetric tilt GBs (STGBs) in perovskite oxides have been investigated by means of atomistic simulations [7–9]. Recent experimental studies [10,11] report that point defects – especially oxygen vacancies – play a significant role for the grain growth behavior in polycrystalline perovskite ceramics of strontium titanate, $SrTiO_3$ (STO), subject to applied electric fields. However, the underlying mechanisms, especially those controlling the redistribution of oxygen vacancies along and across GBs, are not yet fully understood. In order to get a deeper insight into the relationship between grain growth and the presence of oxygen vacancies, classical atomistic calculations of oxygen vacancy formation energies were performed in this study, focusing on the qualitative difference between formation energies at GBs and in bulk regions. With the formation energies, the defect concentrations can be obtained from thermodynamic principles by considering the vibrational and configurational entropy of the defect formation [12–14]. However, in the vicinity of GBs, this is a subject of further research which can build on the work of the present study. Oxygen vacancy formation energies can be used as input for mesoscopic space charge model [15–18] dealing with grain growth under applied electric fields. We selected the two STGBs $\Sigma5$ (310)[001] and $\Sigma5$ (210)[001] because they have been previously investigated by experimental [2,3] and computational studies [7–9], providing a solid foundation for further studies. In the following, they are denoted as $\Sigma5$ (310) and $\Sigma5$ (210) for simplicity.

In this work, we use a rigid-ion model to describe the interatomic interactions, and to obtain oxygen vacancy formation energies. Since oxygen ions are negatively charged, a created vacancy produces a positive charge in an initially neutral simulation cell. In atomistic structure models containing charged layers of ions oriented parallel to GB planes, there is an electric interface dipole moment [19] in the case of GB structures with broken mirror symmetry. Such a dipole moment produces an internal electrostatic potential within the simulation cell, which interacts with the charged vacancy and thereby strongly influences the vacancy formation energy. However, with respect to electrostatics, the supercell approach does not correctly describe the scenario of a macroscopically large crystal, inside which the internal electric field vanishes due to charge compensation effects by various types of point defects in space-charge zones. Finite internal electrostatic potentials in simulation cells have been observed and analyzed in previous studies [20,21] dealing with charged surfaces of 2-dimensional atomistic slab-model systems with periodic boundary conditions. The prototypical scenarios of the charge compensation for charged surfaces were discussed in detail in Ref. [21], and the appropriate electrical boundary condition inside the slab is a vanishing electric field as illustrated in Ref. [20]. Thus, correction methods were formulated to deal with the internal electrostatic potential, which apply an external dipole layer in the vacuum region of a slab-model supercell [20], or which employ an electrostatic surrogate model [21]. However, these approaches were constructed for supercells containing free surfaces and did not take internal interfaces like GBs into consideration. In the present paper, we develop a correction scheme via a 1-dimensional (1D) continuum model based on the surrogate model reported in Ref. [21], to effectively remove artificial electrostatic effects in atomistic simulations of GBs. We apply the developed scheme to both the Σ5 (310) and Σ5 (210) STGBs. Note that such electrostatic artifacts are not expected to exist in more symmetric GB structures which have either a glide-mirror or a screw-rotation symmetry (non-broken mirror symmetries) [22,23], as e.g. those investigated by Genreith-Schriever *et al.* [24]. However, even though one would expect them to be present, internal electric potentials in simulation cells were not addressed in previous classical atomistic-simulation studies similar to the present work, which deal with GBs without the aforementioned symmetric properties [8,25]. We give a possible explanation in our discussion section.

The paper is organized as follows. In Section 2 we describe the details of the calculation method and the simulation model. First, the energetically most favorable configurations of the two considered STGBs in STO are given (Section 2.1). In Section 2.2, we specify the formalism for calculating oxygen vacancy formation energies in the bulk and at a STGB. In Section 2.3, the origin of the electrostatic potentials in the STGB supercell models is described, and the details of the continuum model for correcting for it are specified. We distinguish open boundary conditions in the GB normal direction (Section 2.3.1) and periodic boundary conditions (2.3.2). The formula for the corrected vacancy formation energy is given in Section 2.3.3. In Section 3.1, we apply the correction scheme to the two considered STGBs and demonstrate its validity for both types of boundary conditions. In Section 3.2, we report the resulting profiles of corrected oxygen vacancy formation energies across the boundaries between the two misoriented grains. In the discussion (Section 4), we first compare the application of the correction model for unrelaxed and relaxed GB structures (Section 4.1). Then, the obtained vacancy formation energies are analyzed with respect to the local charge densities, and the differences between the two considered STGBs are discussed (Section 4.2). The electrostatic artifacts in both high-angle and low-angle tilt GBs are illustrated and discussed from a general perspective in Section 4.3. In Section 5, we give a summary and make concluding remarks.

# 2 Methods and model

## 2.1 Atomistic GB structures

This study deals with atomistic supercell models of GBs and point defects in STO. We confine our study to classical molecular-statics (MS) simulations, i.e. to pure structural relaxations, in order to find the equilibrium states of the GBs [26]. For this purpose, we used the program *GULP* (General Utility Lattice Program) [27]. Following Thomas *et al.* [28], we describe the interaction energy between ionic pairs by a rigid-ion model, expressed by a Coulomb-Buckingham potential:

$$U_{ij} = A_{ij} \exp\left(-\frac{r_{ij}}{\rho_{ij}}\right) + \frac{1}{4\pi\epsilon_0}\frac{q_i q_j}{r_{ij}}. \quad (1)$$

Here, $A_{ij}$ and $\rho_{ij}$ are parameters for a pair of ions of types $i$ and $j$, and $r_{ij}$ is the distance between them. $\epsilon_0$ is the vacuum permittivity. The first term describes the short-range repulsive interaction. The second term is the Coulomb interaction between differently charged ions. Partial charges of each ion and parameters of the short-range potential of the considered ion pairs are listed in Table 1. The short-range potential is truncated at a radius of $20$ Å [29]. In the following, Eq.(1) together with these parameters is referred to as "Thomas potential".

Table 1. Parameters of the Thomas potential for the short-range interaction between partial charged ions [28]

| Ion pair | $A$ [eV] | $\rho$ [Å] |
|---|---|---|
| $Sr^{1.84+} - O^{1.40-}$ | 1769.51 | 0.319894 |
| $Ti^{2.36+} - O^{1.40-}$ | 14567.4 | 0.197584 |
| $O^{1.40-} - O^{1.40-}$ | 6249.17 | 0.231472 |

This potential was fitted to reproduce the experimental value $a_{STO} = 3.905$ Å [30,31] of the lattice parameter of STO in the cubic perovskite structure at room temperature. It also is close to the values obtained by density functional theory (DFT) calculations, namely $3.845$ Å obtained by using the local density approximation (LDA) functional [32,33], and $3.942$ Å by applying the generalized gradient approximation (GGA) functional [7,9]. Note however, that the DFT values correspond to zero temperature, whereas the values from experiments and the pair potential refer to room temperature. In addition to reproducing bulk properties, the Thomas potential was verified to be well suitable for describing GBs in STO [29,34] by comparing the atomic structures of some fundamental GBs in this material to results obtained by DFT calculations and TEM experiments. Dislocations [35] and ordered defect configurations [28,36] in STO were also studied with the Thomas potential in the past, demonstrating that it can be applied to different types of bonding environments of the rigid ions. A vacancy also leads to a different neighbor structure of the surrounding ions, and the crystal structures containing a vacancy in the bulk or at the GBs were shown to remain stable upon relaxation.

We therefore apply the Thomas potential to describe oxygen vacancies in the bulk and at the GB as well, as it was done previously by Schie *et al*., who studied oxygen vacancy diffusion in STO by Molecular Dynamics simulations [37]. However, absolute defect energies are not necessarily in good agreement with DFT or experimental values, which was shown for GB energies in STO derived by the Thomas potential by Benedek *et al.* [29]. But the potential was also shown there to reproduce the hierarchy of energies of different GBs correctly. As will be described in Section 2.2, we are not interested in absolute defect energies in this work, but rather in qualitative energy differences of vacancies in bulk and GB environments, which in our opinion can be well described by the Thomas potential.

Applying the method developed in Refs. [7] and [9], we generated supercells of STO containing the symmetric tilt grain boundaries (STGBs) $\Sigma5$ (310) and $\Sigma5$ (210). Here and in the following we will choose the orientation of our model structures such that the $y$-axis (lattice parameter $b$) is perpendicular to the GB plane. Considering the coincidence site lattice (CSL) [38], the cell parameters in the GB plane, $a_{\mathrm{cell}}$ and $c_{\mathrm{cell}}$, are each set to one CSL elementary-cell length [$a_{\mathrm{cell}} = \sqrt{10}a_{\mathrm{STO}}$ for $\Sigma5$ (310) and $a_{\mathrm{cell}} = \sqrt{5}a_{\mathrm{STO}}$ for $\Sigma5$ (210)], and to one lattice constant of STO ($c_{\mathrm{cell}} = a_{\mathrm{STO}}$ for both GBs), respectively. The cell parameters in the GB normal direction ($b_{\mathrm{cell}}$) are taken initially as four times the CSL elementary-cell lengths, and after structural optimization, they are approximately $52$ Å for the STGBs $\Sigma5$ (310) and $36$ Å for $\Sigma5$ (210). This choice of cell lengths ensures that the bulk regions are sufficiently large in order to minimize finite size errors, which was confirmed by preliminary convergence tests with respect to the GB energy $\gamma$ (as defined in Ref. [29]). In total, the STGB supercells contain 200 atoms for $\Sigma5$ (310) and 100 atoms for $\Sigma5$ (210). Since periodic boundary conditions were applied in the generation and optimization procedure of the GB structures, each supercell contains two identical GBs.

In order to identify the stable, i.e. energetically most favorable GB configurations, rigid-body translations (RBT) and structural relaxations were carried out. Following the methodology outlined in Refs. [8] and [9], we performed stepwise rigid-body translations (RBT) of one grain with respect to the other in all three directions. We determined the most stable GB configurations by comparing the GB energies $\gamma$ after structural relaxation at each RBT step.

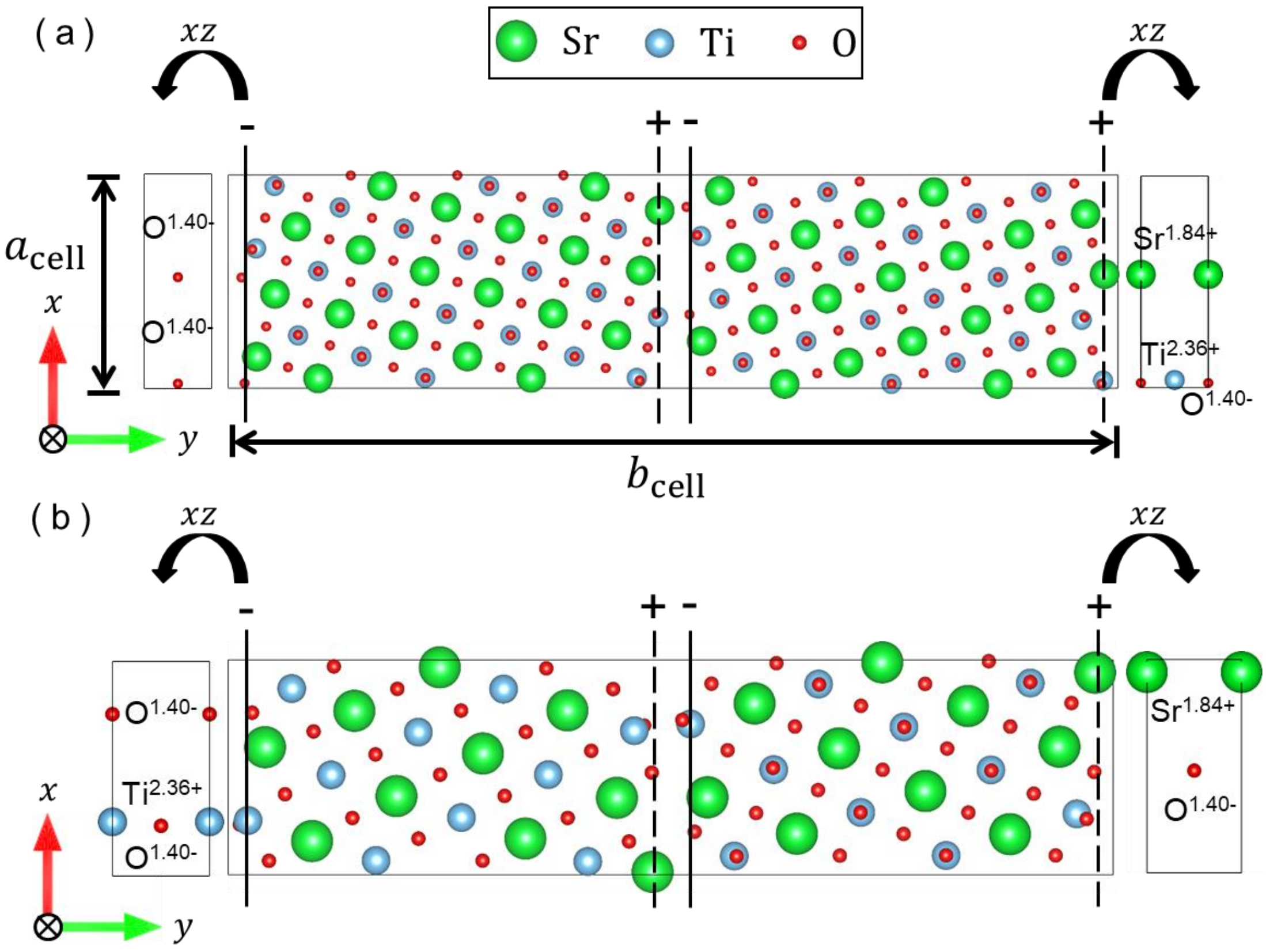

Figure 1. The relaxed configurations viewed from [001] direction: (a) STGB Σ5 (310), and (b) STGB Σ5 (210). The cell parameters $a_{\mathrm{cell}}$ and $b_{\mathrm{cell}}$ denote the supercell dimensions in $x$- and $y$-directions, respectively. The third cell parameter, $c_{\mathrm{cell}}$, denoting the cell length in $z$-direction is not shown. The $xz$-planes of each configuration are either positively or negatively charged. The ionic compositions leading to those charges are exemplary shown for the GB planes at the two ends of the supercells and marked by dashed (positive charge) and solid lines (negative charge) in the structure models.

The optimized atomic structures are displayed in Figure 1, for which we obtained $\gamma = 1.81\ \mathrm{J/m^2}$ for STGB Σ5 (310) and $\gamma = 1.54\ \mathrm{J/m^2}$ for Σ5 (210). In terms of structural units, these structures are in good agreement to those obtained by DFT-GGA studies in Ref. [7] (Figure 5(c), $\gamma = 1.02\ \mathrm{J/m^2}$) and Ref. [9] (Figure 4(b), $\gamma = 0.98\ \mathrm{J/m^2}$), for STGB Σ5 (310) and Σ5 (210), respectively. In the following, we denote our two optimized structures as "relaxed configurations" to distinguish them from their corresponding initial structures before structural relaxation, which are named "unrelaxed configurations".

Considering the $xz$-planes in the supercells, the unrelaxed structures are composed of repeated units of two types of atomic layers: a unit containing a Sr-Ti-O and an O-O plane in the case of Σ5 (310), and a unit containing a Sr-O and a Ti-O-O plane in the case of Σ5 (210). The configurations of these units largely remain after structural relaxation, merely the ions close to the GBs deviate slightly from their positions at the lattice planes of the unrelaxed structures. This can be seen in Figure 1, where we also display the different layers in the $xz$-plane at the two ends of the supercells. Considering the partial charges of the ionic species used in the Thomas potential (see Table 1), the Sr-Ti-O plane and the O-O plane in the $\Sigma5$ (310) cell are positively and negatively charged with $2.80\ \mathrm{e}$, respectively. In the $\Sigma5$ (210) cell, the Sr-O plane and Ti-O-O plane are positively and negatively charged by $0.44\ \mathrm{e}$, respectively. Hence, the obtained configurations of both considered GB structures can be represented by periodically repeated units consisting of two oppositely charged planes. Note however, that in the case of $\Sigma5$ (210), the planes would be charge neutral if the formal ionic charges of Sr, Ti and O ($+2\ \mathrm{e}$, $+4\ \mathrm{e}$ and $-2\ \mathrm{e}$) were taken instead of the Thomas charges (*cf.* Section 4.3).

## 2.2 Oxygen vacancy formation energy

Oxygen ions in the supercell were separately removed and the respective vacancy formation energy was calculated. The formation energy $E_\mathrm{f}$ of a vacancy in the rigid-ion model can be expressed as [8]:

$$E_\mathrm{f} = E_\mathrm{tot} - E_\mathrm{tot}^{(0)} + E_\infty + E_\mathrm{corr}, \tag{2}$$

where $E_\mathrm{tot}$ is the total lattice energy of the supercell containing the defect. If the vacancy is charged, $E_\mathrm{tot}$ can be calculated by introducing a neutralizing uniform background charge density [27]. $E_\mathrm{tot}^{(0)}$ denotes the total lattice energy of the supercell without defect. $E_\infty$ is the energy of the removed neutral atom or charged ion being placed isolated at infinite separation from the lattice. In DFT calculations of charged point defects, one needs to consider a correction term $E_\mathrm{corr}$, which generally includes a periodic image charge correction and a potential alignment [39]. In the classical MS calculations of this work, we denote by $E_\mathrm{corr}$ the correction of the energy from the interaction of the charged oxygen vacancy defect with the electrostatic potential stemming from the charged lattice planes in the supercells containing GBs. This will be explained in detail in the following section. In addition, the interaction of an oxygen vacancy with its own periodic images needs to be corrected. We confirmed by preliminary calculations that the Coulomb interaction energy between periodic images of the oxygen vacancy can be decreased to below $0.1$ eV by extending the cell dimension in $z$-direction to three units ($c_\mathrm{cell} = 3a_\mathrm{STO}$) for both STGBs.

The central two terms on the right-hand side of Equation (2) are independent of the position of the oxygen vacancy in the supercell. Because we are primarily interested in the influence of GBs on oxygen vacancies, we introduce the formation energy difference $\Delta E_\mathrm{f}(y)$ of an oxygen vacancy at a position $y$ in the cell with respect to the formation energy of an oxygen vacancy located at a reference point $y_\mathrm{B}^\mathrm{ref.}$ inside one of the two adjacent grains:

$$\Delta E_\mathrm{f}(y) = E_\mathrm{f}(y) - E_\mathrm{f}(y_\mathrm{B}^\mathrm{ref.}). \tag{3}$$

By inserting Equation (2) we can reformulate this expression:

$$\Delta E_\mathrm{f}(y) = E_\mathrm{tot}(y) - E_\mathrm{tot}\left(y_\mathrm{B}^\mathrm{ref.}\right) + E_\mathrm{corr}(y) - E_\mathrm{corr}\left(y_\mathrm{B}^\mathrm{ref.}\right). \tag{4}$$

## 2.3 Continuum model for correcting electrostatic artifacts

As described in Section 2.1, the structural configurations of both considered STGBs cells consist of periodically repeated units of two oppositely charged planes parallel to the GB ($xz$-) plane. Such a repeated-units structure is schematically sketched in Figure 2, containing the two bulk regions of the grains, which are separated by a GB region. Note that the GB separation is generally larger than the distances between the planes in the bulk regions after GB relaxation because of a positive GB excess volume.

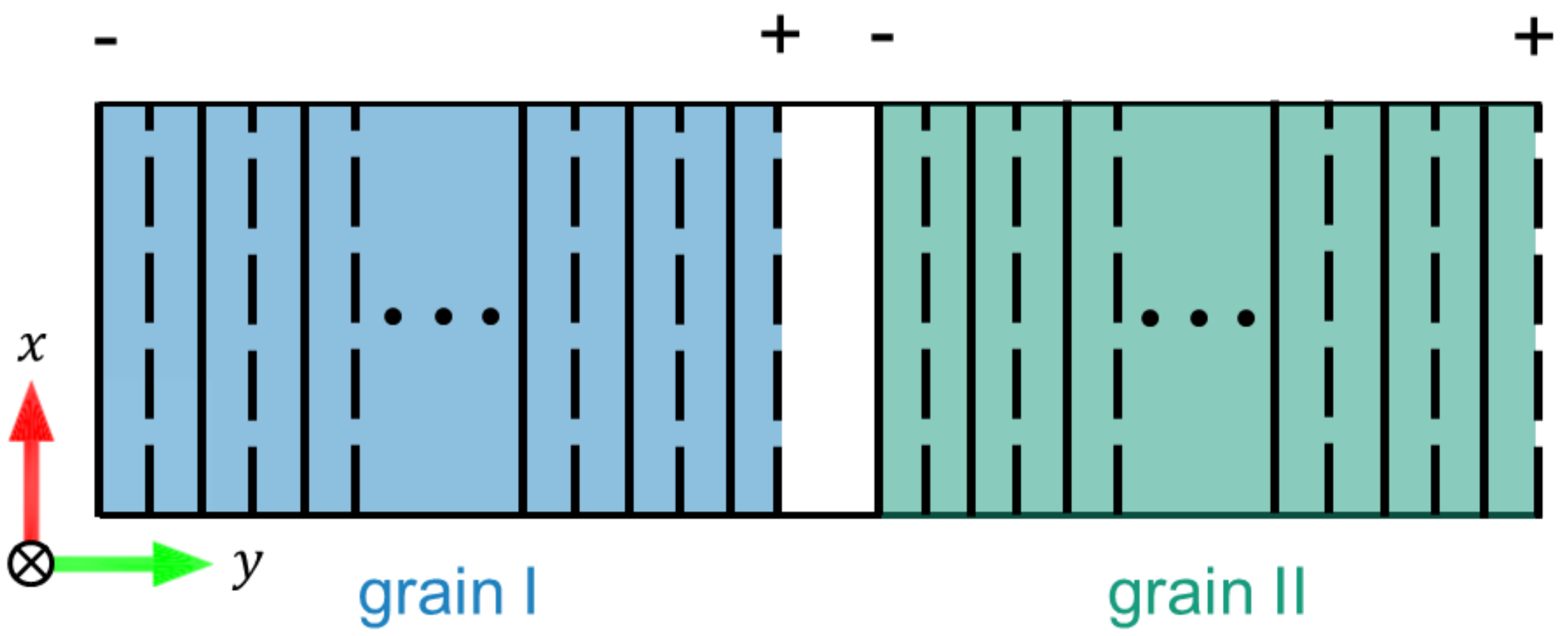

Figure 2. A sketch of a general STGB supercell. Both open and periodic boundary conditions can be applied in the direction perpendicular to the GB plane (the $y$-direction). Differently colored regions mark the generally differently oriented grains. As introduced in Figure 1, the vertical dashed and solid lines indicate positively and negatively charged planes, respectively.

Here, vertical solid lines denote negatively charged planes while vertical dashed lines indicate positively charged planes (*cf.* Figure 1). Using open boundary conditions in $y$-direction, such a stacking sequences of alternatingly charged planes produces a dipole moment perpendicular to the surface, as it is qualitatively described in Ref. [19]. This dipole moment produces an internal electrostatic potential, which interacts with charged species such as positively charged oxygen vacancies, and thereby strongly influences the vacancy formation energy. Such electrostatic effects have been encountered and analyzed in previous studies [20,21] dealing with charged surfaces in 2-dimensional surface-slab systems. They are reported as simulation artifacts of the approach, because according to classical electrostatics, as explained in Refs. [20,21], the internal electric field in a defect-free region of a crystal is zero.

Thus, correction methods were proposed to remove the internal potential, such as applying an external dipole layer or using an electrostatic surrogate model [20,21]. However, these approaches were constructed for systems containing free surfaces only and did not take internal GBs into consideration. For GBs with broken mirror symmetry, such as the STGBs $\Sigma5$ (210) and of $\Sigma5$ (310) treated in this work, an additional dipole moment from the internal interface needs to be considered. Such a dipole moment does not appear in GB supercells with mirror symmetry, because there, the dipole moments from the two mirror symmetric grains cancel each other out. To effectively remove the simulation artifact of an internal electrostatic potential in our study, we first investigate the supercell with *open* boundary conditions in the direction perpendicular to the GB plane (the $y$-direction). Using such boundary conditions, the complications arising by periodic images of the GB planes due to *periodic* boundary conditions in the GB normal direction can be avoided at first. The electrostatic potential for open boundary conditions is derived via a 1-dimensional (1D) continuum model. Based on this model, the electrostatic potential in a supercell with periodic boundary conditions is derived in the next step.

For ease of readability, we distinguish the two terms “surface” and “interface” in the following discussions. We use “surface” when we refer to an interface between a bulk material on one side and vacuum on the other side, whereas an “interface” implies bulk materials on both sides, as in the region of a GB.

### 2.3.1 The electrostatic potential for open boundary conditions

To derive the electrostatic potential inside of a supercell containing a GB, we first consider the electrostatic potential arising from an isolated, homogeneously charged plane with an area $A$ and total charge $q$, corresponding to a charge density $q/A$. The normal of the plane is oriented in $y$-direction, and if $A$ becomes infinitely large, we can interpret the problem as a one-dimensional scenario. Supposing the plane is located at $y = y_q$, the electrostatic potential $V_q$ obeys the one-dimensional Poisson equation:

$$\frac{d^2V_q}{dy^2} = -\frac{q}{A\epsilon_0}\delta(y - y_q), \tag{5}$$

with $\delta$ representing the Dirac delta function. The general solution can be found as:

$$V_q(y) = -\frac{q}{2A\epsilon_0}(y - y_q) \cdot [H(y - y_q) - H(y_q - y)] + C_1 y + C_2, \tag{6}$$

where $H$ denotes the Heaviside step function, and $C_1$ and $C_2$ are constants of integration, which need to be determined based on the boundary conditions. Taking the negative derivative yields an electric field $E$ in the following form [19]:

$$E = \begin{cases} -\frac{q}{2A\epsilon_0} - C_1, \ y < y_q \\ \frac{q}{2A\epsilon_0} - C_1, \ y > y_q \end{cases}, \tag{7}$$

so $C_1$ represents an additional, constant electric field.

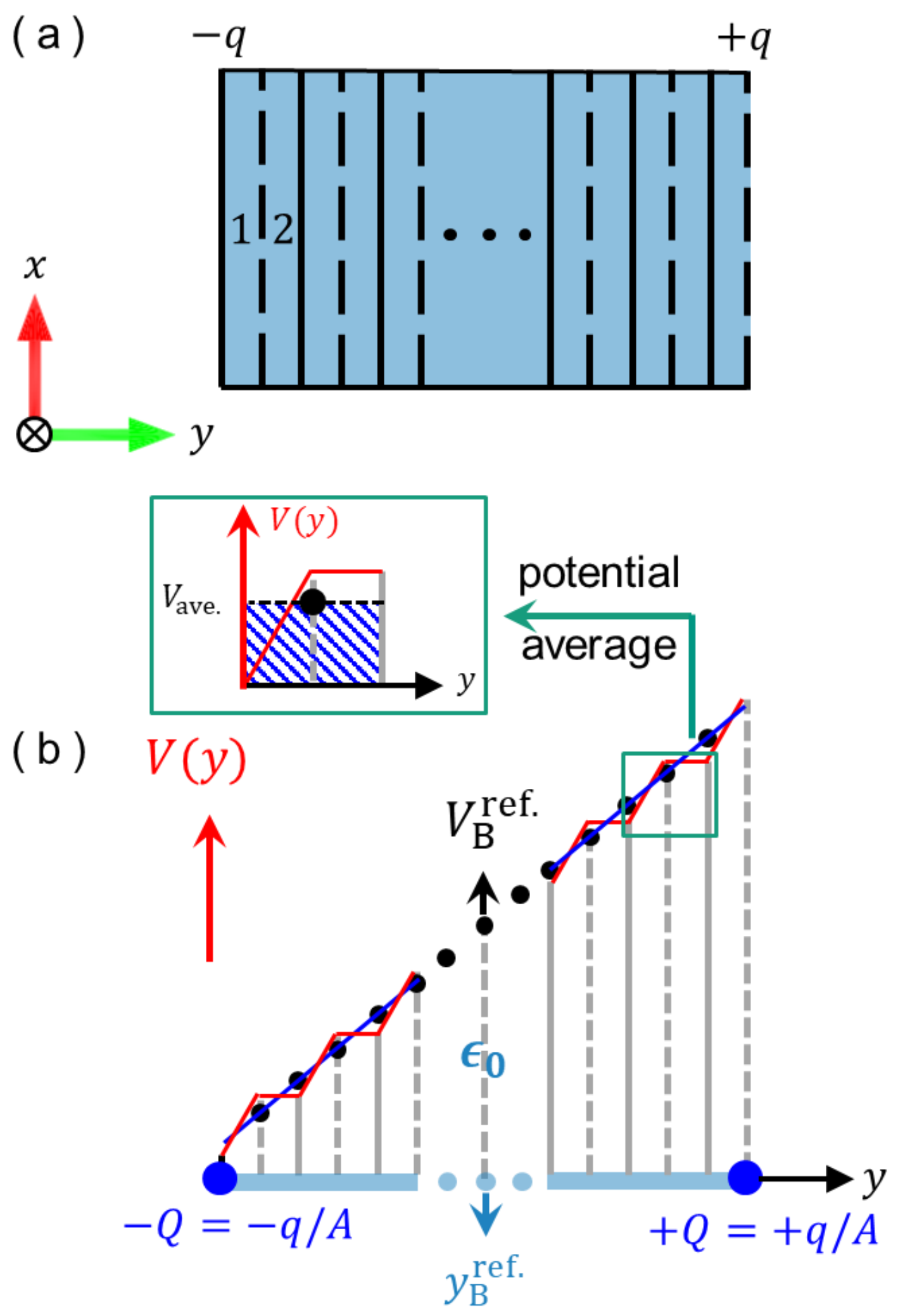

Figure 3. (a) A sketch of a general supercell with open boundary conditions in $y$-direction and equidistant, alternatingly charged planes; (b) the resulting schematic electrostatic potential (red line). The inset shows the method of determining the averaged potential at the position of a plane (details in the text). The averaged potential at each plane is marked with a black point, and the connection of the black points leads to the blue line, representing a linear function of $y$. A reference point $y_{\mathrm{B}}^{\mathrm{ref.}}$ in the bulk region of a grain is selected, with the potential at this point $V_{\mathrm{B}}^{\mathrm{ref.}}$. Such an averaged potential (blue line) can be modelled by two oppositely charged point charges $\pm Q$ located at the outermost positions of the supercell in $y$-direction, as shown by the blue dots.

Next, we consider a general interface supercell like the one sketched in Figure 2, with open boundary conditions in $y$-direction. The total electrostatic potential $V^{\mathrm{tot}}(y)$ can be calculated by summing over the contributions from all charged planes [*cf.* Equation (6)]:

$$V^{\mathrm{tot}}(y) = -\sum_i \frac{q_i}{2A\epsilon_0}(y - y_i) \cdot [H(y - y_i) - H(y_i - y)] + C_1 y + C_2. \quad (8)$$

In the case of open boundary conditions without an additional, external electric field, we can set $C_1 = 0$. A particular case is sketched in Figure 3(a) with equidistant, alternatingly charged planes (with planar averaged charge densities $\pm q/A$). An electric field of the same magnitude alternatingly appears or vanishes in the regions between the planes. In the evenly numbered regions [e.g. the region labeled 2 in Figure 3(a)], there is no electric field because of the same number of negatively and positively charged planes on the left side and on the right side, respectively. In each of the oddly numbered regions, an electric field originating from the two outermost, unpaired surface planes is left in any neutral supercell. The corresponding electrostatic potential $V^{\mathrm{tot}}(y)$ is schematically sketched by the red stage-like curve in Figure 3(b).

We now define the potential at each charged plane as the averaged potential between the two adjacent planes. As shown in the inset of Figure 3(b), the averaged potential (indicated by the black point) is determined such that the area under red lines is of the same size as the shadowed blue region. This concept was introduced by Harrison *et al.* [40] to understand the electrostatics at interfaces in polar semiconductors containing alternatingly charged planes.

Since the planes are regularly spaced, the average potential inside the bulk regions (straight blue line) corresponds to an average electric field of half the strength compared to the one within the oddly numbered regions. Extrapolating it to the two surface planes with charge densities $-q/A$ and $+q/A$, the averaged potential can be macroscopically modelled by a dipole with two point charges $\pm Q = \pm q/A$, located at the two charged surfaces, as shown by the blue dots in Figure 3(b).

Considering the factor 1/2 according to this procedure, the averaged electrostatic potential within the supercell can be derived by applying Equation (6) to a system with two oppositely charged planes:

$$V_{\mathrm{ave.}}(y) = V_{\mathrm{B}}^{\mathrm{ref.}} + \frac{q}{2A\epsilon_0} \cdot \left(y - y_{\mathrm{B}}^{\mathrm{ref.}}\right). \tag{9}$$

Here, we introduced the reference point $y_{\mathrm{B}}^{\mathrm{ref.}}$ lying in the bulk region (B) of a grain, with the potential at this point being $V_{\mathrm{B}}^{\mathrm{ref.}}$.

Finally, the model can be applied to a cell containing a STGB, as sketched in Figure 4. As the central GB region separates the supercell into two differently oriented regions, four point charges $Q_i$ $(i = 1, 2, 3, 4)$ (indicated by the black dots) are introduced representing the averaged surface charges $\pm q/A$. Each point charge is placed at a position $y_i$, which is the position of the respective charged plane.

Again, we average the stage-like potential (red curve) originating from all charged planes at the positions of these planes (black points), which yields the blue line. Choosing the reference point arbitrarily somewhere in region I, the averaged potential $V_{\mathrm{I}}(y)$ in this region is given by Equation (9). The corresponding averaged potential $V_{\mathrm{II}}(y)$ in grain II must have the same slope, but since the grain boundary excess separation ($\delta_{\mathrm{GB}}$) is different (usually larger) than the separation of lattice planes in the grains ($d$), there is an offset between the lines representing $V_{\mathrm{I}}(y)$ and $V_{\mathrm{II}}(y)$. In order to match them, $V_{\mathrm{I}}(y)$ is extrapolated to $\tilde{y}_2 = y_2 + \Delta y$ and $V_{\mathrm{II}}(y)$ to $\tilde{y}_3 = y_3 - \Delta y$, such that the potential is constant between these points, i.e. $V_{\mathrm{I}}(\tilde{y}_2) = V_{\mathrm{II}}(\tilde{y}_3)$ (see Figure 4). A constant potential in this intermediate region is reasonable since the two neighboring grains are charge neutral. This corresponds to the solution of the Poisson equation [Equation (6)] for a system of four point charges. Thus, the averaged potential reads:

$$V_{\text{ave.}}(y) = V_{\text{B}}^{\text{ref.}} + \frac{q}{2A\epsilon_0}\left(y - y_{\text{B}}^{\text{ref.}}\right) \text{ for } y_1 \leq y \leq \tilde{y}_2, \quad \text{(10-a)}$$

$$V_{\text{ave.}}(y) = V_{\text{B}}^{\text{ref.}} + \frac{q}{2A\epsilon_0}\left(\tilde{y}_2 - y_{\text{B}}^{\text{ref.}}\right) \text{ for } \tilde{y}_2 < y < \tilde{y}_3, \quad \text{(10-b)}$$

$$V_{\text{ave.}}(y) = V_{\text{B}}^{\text{ref.}} + \frac{q}{2A\epsilon_0}(2\Delta\text{y} - \delta_{\text{GB}}) + \frac{q}{2A\epsilon_0}\left(y - y_{\text{B}}^{\text{ref.}}\right) \text{ for } \tilde{y}_3 \leq y \leq y_4. \quad \text{(10-c)}$$

The GB excess separation is given by $\delta_{\text{GB}} = y_3 - y_2$. The value of the parameter $\Delta$y will be discussed below when the model is applied to an actual GB scenario.

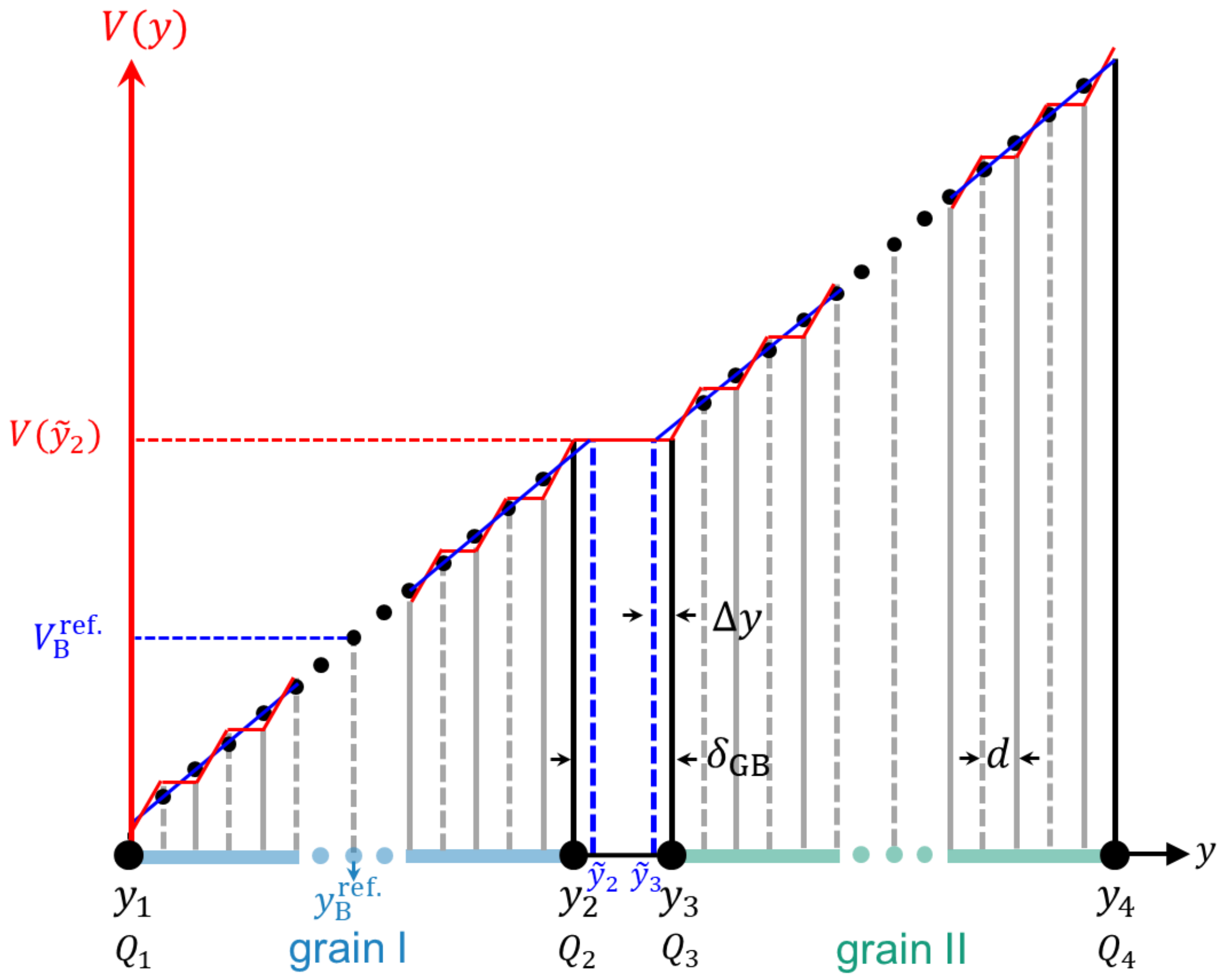

Figure 4. Schematic sketch of the electrostatic potential in the general 1D continuum model of a supercell containing a STGB with differently charged termination planes. An arbitrarily chosen reference point $y_{\text{B}}^{\text{ref.}}$ within grain I (here: the midpoint with respect to $y$) is selected, at which the averaged potential is defined as $V_{\text{B}}^{\text{ref.}}$. Lattice planes in bulk regions are equally spaced with a distance $d$, and the GB separation $\delta_{\text{GB}}$ is usually larger than $d$. Extrapolating the straight blue lines of the average potentials within both bulk regions, they intercept with the red stage-like curve at position $\tilde{y}_2$ and $\tilde{y}_3$, which deviate from $y_2$ and $y_3$ by $\pm\Delta$y. Such an averaged potential can be modelled by considering four point charges $Q_1$ to $Q_4$ along the $y$-axis, placed at the positions of the outer layers of each grain (large black dots). Here, $Q_1 = Q_3 = -\frac{q}{A}$ and $Q_2 = Q_4 = +\frac{q}{A}$, where $q$ represents the charge of the surface plane with area of $A$.

### 2.3.2 The electrostatic potential for periodic boundary conditions

In the previous section, we have obtained the formula for the internal electrostatic potential induced by charged lattice planes when *open* boundary conditions are applied to a STGB supercell as the one sketched in Figure 2. The electrostatic potential monotonically increases along the whole length of the cell in $y$-direction. However, in the case of *periodic* boundary conditions in this direction, the potential has to be periodic, too. From the viewpoint of electrostatics, both bulk grain regions I and II are identical, since they have equivalent environments of charged GB termination planes. Therefore, the potential at equivalent positions in the two bulk regions has to be identical. Such a potential can be obtained by subtracting a linear function, which connects the potentials at two equivalent positions in the two regions, from the non-periodic potential for open boundary conditions [Equation (10)], as visualized in Figure 5. This procedure corresponds to the dipole correction [41], which has been widely used in order to treat the internal electric field induced by a surface dipole in surface geometries. The total internal electric field in the direction perpendicular to the GB for the *open* boundary conditions in our study contains both influences from the outer surfaces and the central interface. After removing the surface contribution by subtracting a linear potential:

$$V_{\mathrm{surf}}(y) = V_{\mathrm{B}}^{\mathrm{ref.}} + \frac{q}{2A\epsilon_0} \cdot \frac{L+2\Delta y}{L+\delta_{GB}} \cdot \left(y - y_{\mathrm{B}}^{\mathrm{ref.}}\right) \text{ for } y_1 \leq y \leq y_4, \quad (11)$$

only the potential caused by the interface remains. As described in Section 2.3.1 [Equation (6)], subtracting a linear potential ($C_1 y$) still satisfies the Poisson equation, which has to be done here in order to fulfill the periodic boundary conditions. In Equation (11), $L = y_4 - y_3 = y_2 - y_1$ denotes the length of each of the two grain regions, and $\delta_{\mathrm{GB}} = y_3 - y_2$ as defined above. Note that in the case of periodic boundary conditions, the cell length in $y$-direction is given by $b_{\mathrm{cell}} = 2L + 2\delta_{\mathrm{GB}}$. Subtracting Equation (11) from Equation (10), the remaining averaged potential reads:

$$V_{\mathrm{ave.}}(y) = \frac{q}{2A\epsilon_0} \frac{\delta_{\mathrm{GB}} - 2\Delta y}{L+\delta_{\mathrm{GB}}} \left(y - y_{\mathrm{B}}^{\mathrm{ref.}}\right) \text{ for } \tilde{y}_1 \leq y \leq \tilde{y}_2, \quad (12\text{-a})$$

and

$$V_{\mathrm{ave.}}(y) = \frac{q}{2A\epsilon_0} \frac{\delta_{\mathrm{GB}} - 2\Delta y}{L+\delta_{\mathrm{GB}}} \left(y - y_{\mathrm{B}}^{\mathrm{ref.}}\right) + \frac{q}{2A\epsilon_0} (2\Delta \mathrm{y} - \delta_{\mathrm{GB}}) \text{ for } \tilde{y}_3 \leq y \leq \tilde{y}_4. \quad (12\text{-b})$$

Note that due to the periodic boundary conditions, we have to introduce $\tilde{y}_1 = y_1 - \Delta y$ and $\tilde{y}_4 = y_4 + \Delta y$ in the same way as explained in Section 2.3.1 for $\tilde{y}_2$ and $\tilde{y}_3$. Between $\tilde{y}_2$ and $\tilde{y}_3$ as well as between $\tilde{y}_4$ and $\tilde{y}_1$, the potential linearly decreases, leading to a saw-tooth profile as sketched in Figure 5. Since the potential profile in the interface regions $\tilde{y}_2 \leq y \leq \tilde{y}_3$ and $\tilde{y}_1 \geq y \geq \tilde{y}_4$ is not relevant for the following defect calculations (there is no oxygen site), the potential function is not explicitly given here, but can easily be derived.

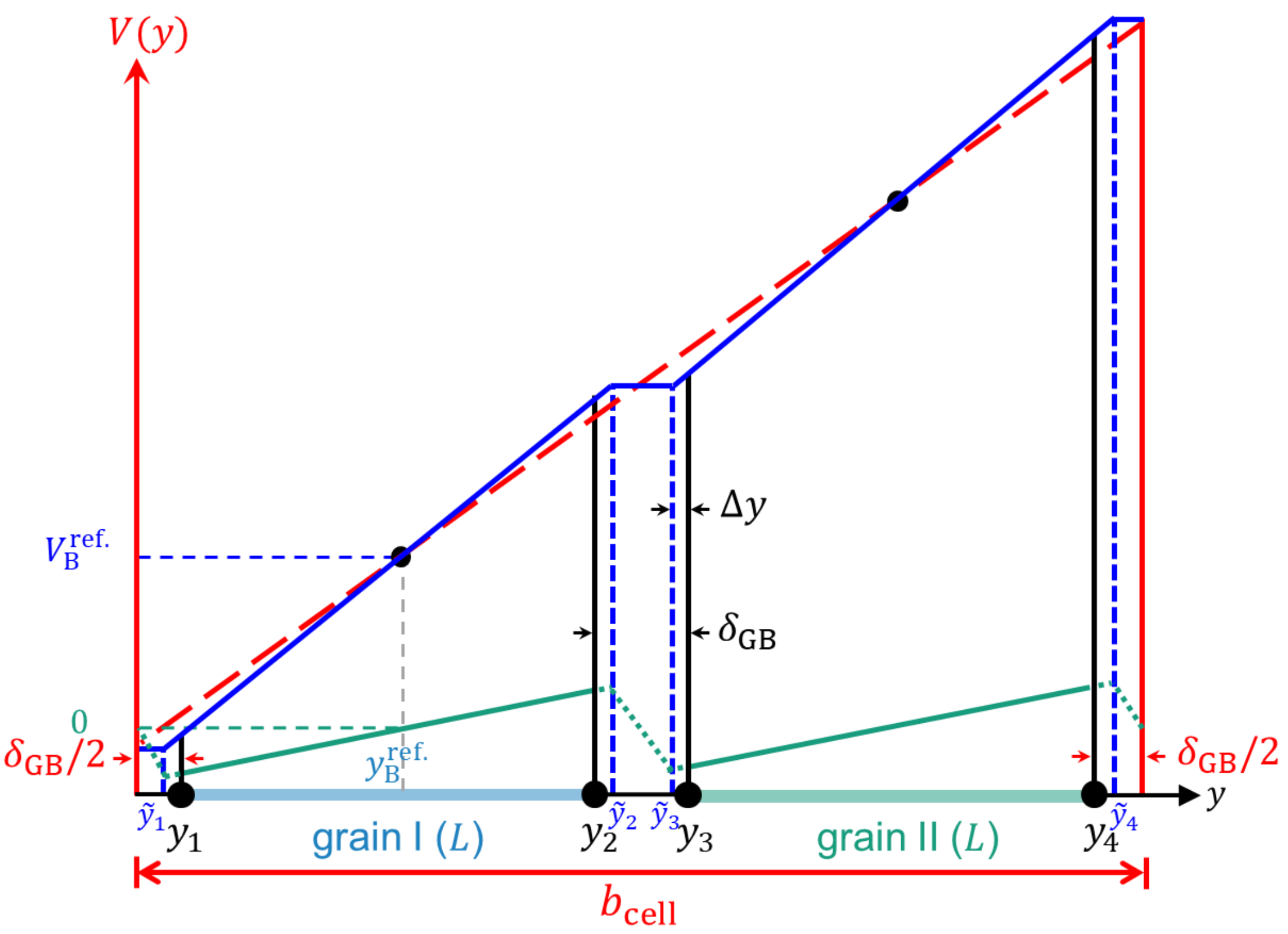

Figure 5. Derivation of the electrostatic potential in a supercell with periodic boundary conditions. The potential for open boundary conditions in $y$-direction is given by the blue line. The red dashed line, connecting the reference point in region I with the identical point in region II (here e.g. the two midpoints of the bulks regions) represents the electrostatic potential of the surface dipole. The difference between the blue and the red potential lines is given in green, with amplified slopes (by a factor of 2) for a better visualization. The potentials in grain I [grain II] are extrapolated to the points $\tilde{y}_1$ and $\tilde{y}_2$ [$\tilde{y}_3$ and $\tilde{y}_4$] inside of the GB regions. $L$ denotes the length of each of the two bulk grain regions.

### 2.3.3 The correction energy

After having derived the formulae of the averaged electrostatic potentials $V_{ave.}(y)$ for supercells with both, open and periodic boundary conditions along the $y$-direction perpendicular to the interface, the correction energy of a defect with charge $q^d$ at a position $y$ in the supercell $E_{corr}(y)$ is given by:

$$E_{corr}(y) = -q^d V_{ave.}(y). \tag{13}$$

The minus sign indicates the removal of the electrostatic effects by subtracting the corresponding energy. Note that, in general, the accurate electrostatic potential should be the potential generated by all of the ions *excluding* the removed O-ion at position $y$ [42]. The term “electrostatic potential” used in all of the above derivations however refers to the total electrostatic potential from summing over *all* ions in the investigated supercells. The electrostatic self-energy of an ion is independent of its position, and would therefore be cancelled if Equation (13) is inserted into Equation (4). Hence, we obtain the formula for the corrected defect formation energy difference between an arbitrary point $y$ in the supercell and the reference point $y_{\mathrm{B}}^{\mathrm{ref.}}$ in the bulk as:

$$\Delta E_{\mathrm{f}} = E_{\mathrm{tot}}(y) - E_{\mathrm{tot}}\left(y_{\mathrm{B}}^{\mathrm{ref.}}\right) - q^{\mathrm{d}}[\, V_{\mathrm{ave.}}(y) - V_{\mathrm{B}}^{\mathrm{ref.}}]. \quad (14)$$

Note that if the same reference point $y_{\mathrm{B}}^{\mathrm{ref.}}$ is chosen for Equation (14) and for the potentials $V_{\mathrm{ave.}}(y)$ [Equation (10) and (12)], the term $V_{\mathrm{B}}^{\mathrm{ref.}}$ cancels out of Equation (14). In the continuum model developed so far, there are still two parameters to be determined: the surface charge density $q/A$ and the position deviation $\Delta$y. This determination will be done in Section 3.1 for the cases of oxygen vacancies in supercells containing the two grain boundaries $\Sigma$5 (310) and $\Sigma$5 (210).

# 3 Results

## 3.1 Correction of the electrostatic artifacts

At the example of a supercell containing the STGB Σ5 (310), the correction methodology is demonstrated for the calculation of formation energies of positively charged oxygen vacancies ($V_{\mathrm{O}}^{+1.4}$). For both open (Section 3.1.1) and periodic boundary conditions (Section 3.1.2), unrelaxed GB supercells are analyzed first, and then the effect of structural relaxation is considered for the case of periodic boundary conditions. Note, that for each of these cases, the STGB structures were not relaxed any more *after* an oxygen ion was removed. It is shown at the end of Section 3.1.2 that a relaxation of the system with vacancy leads to a considerable decrease of the total energy of the system, but it has no influence on the electrostatic potential and the application of the correction scheme.

### 3.1.1 The correction for open boundary conditions

In order to calculate the oxygen vacancy formation energies as described in Section 2.2, vacancies were first generated separately on all possible oxygen sites of an unrelaxed supercell containing a STGB Σ5 (310). Open boundary conditions were applied in the direction perpendicular to the GB, and periodic boundary conditions in the directions parallel to the GB. The formation energies were calculated and referenced to the formation energy of a defect located at a position $y_{\mathrm{B}}^{\mathrm{ref.}}$ in the bulk grain region. The values obtained without applying the electrostatic correction are shown in Figure 6 (labeled as “simulation data”). The effect of the electrostatic potential is clearly visible by the strong increase of the values by about $350$ eV across the supercell.

In order to correct this apparent artifact, we use Equation (14) and the potential function given by Equation (10). We use the same reference point $y_{\mathrm{B}}^{\mathrm{ref.}}$ for the formation energy and for the potential.The defect charge is the charge of the oxygen vacancy ($q^{\mathrm{V_O}} = +1.40\ e$).

As explained in Section 2.1, the unrelaxed supercell is composed of positively charged Sr-Ti-O planes ($+q$) and negatively charged O-O planes ($-q$). Considering the partial charges of the ionic species in the Thomas potential, the value of $q$ is equal to $2.80\ e$ on one lattice plane of area $A = 48.22\ \mathrm{Å}^2$. The positions $y_i$ of the four point charges in the 1-D model are specified by the positions of the charged surface ($y_1, y_4$) and interface planes ($y_2, y_3$). Hence, the GB separation is $\delta_{\mathrm{GB}} = y_3 - y_2$. In the unrelaxed configurations, the lattice planes in the two bulk regions are equally spaced by a distance $d = 0.618\ \mathrm{Å}$ , which leads to a deviation parameter $\Delta y = d/2$.

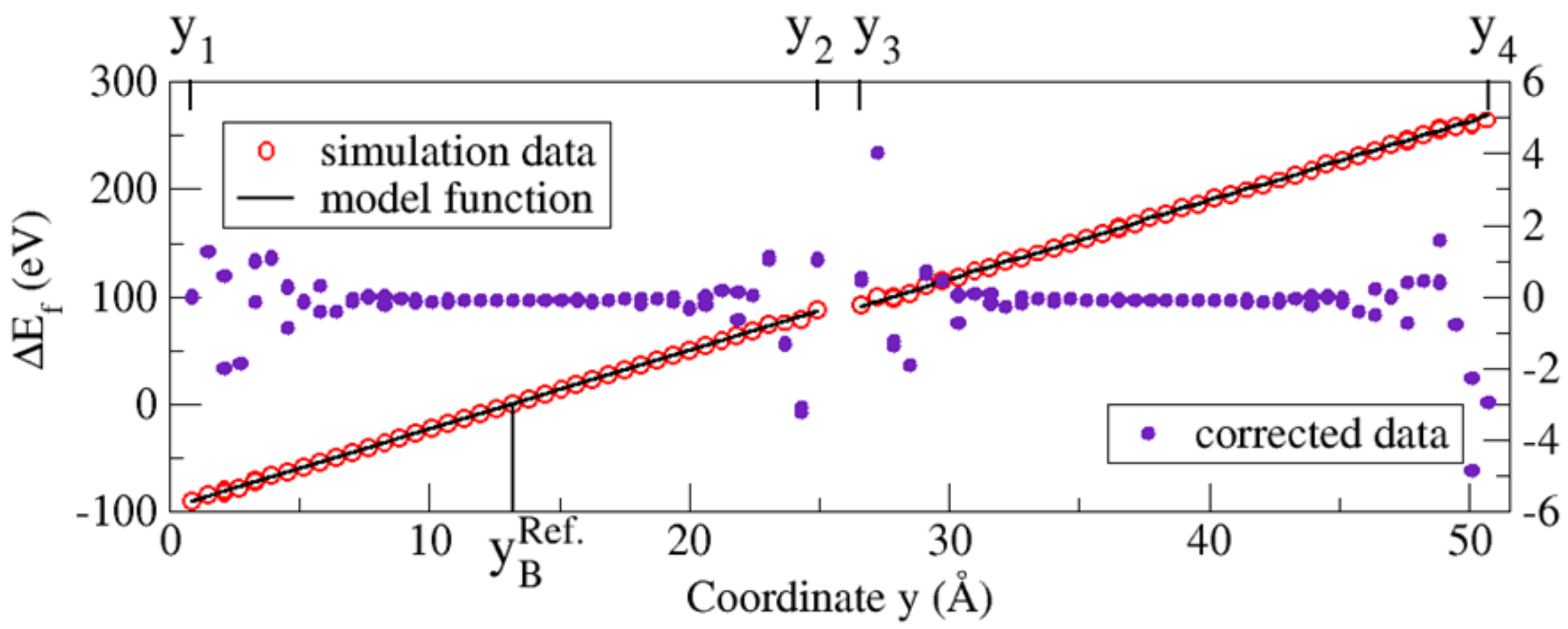

Figure 6. The relative formation energy of oxygen vacancies ($\Delta E_{\mathrm{f}}$) for the unrelaxed configuration of a supercell containing a STGB Σ5 (310) with open boundary condition in the GB normal ($y$) direction. The positions of the charged planes ($y_i$) are labeled on the top of the graph. Potentials are referenced to the value at the position $y_{\mathrm{B}}^{\mathrm{ref.}}$. The simulation data calculated with GULP [$E_{\mathrm{tot}}(y) - E_{\mathrm{tot}}(y_{\mathrm{B}}^{\mathrm{ref.}})$], indicated by red circles (left $y$-axis), match the line of the potential model, marked by the black line. The values corrected for the internal electrostatic potential (purple points) are plotted with respect to the rescaled right $y$-axis for a better visibility.

Using these parameters, a successful correction was achieved using the 1-D continuum model, as shown in Figure 6. In the bulk regions, the uncorrected simulation data points deviate from those of the model function on the order of $0.01$ eV, which confirms the validity of our correction approach.

### 3.1.2 The correction for periodic boundary conditions

In the case of periodic boundary conditions in the direction perpendicular to the GB, the model function to correct for the electrostatic potential in the bulk regions of the unrelaxed Σ5 (310) supercell is obtained by combining Equations (12) and (14). Here, the bulk grain length $L = y_4 - y_3 = y_2 - y_1 = 24.08$ Å is needed as an additional parameter. The values of the other parameters are the same as those used in Section 3.1.1.

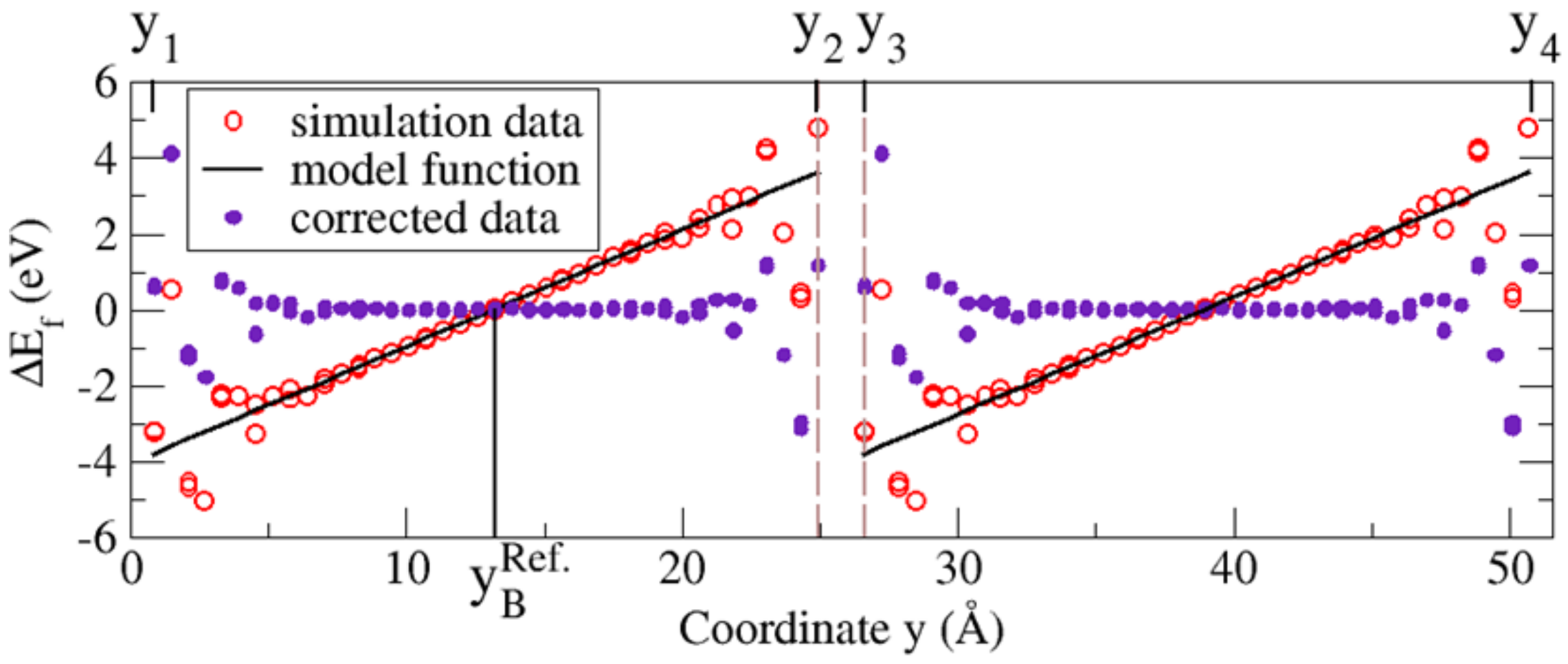

Figure 7. The relative formation energy of oxygen vacancies ($\Delta E_{\mathrm{f}}$) for the unrelaxed configuration of a cell containing a STGB $\Sigma$5 (310), with periodic boundary condition in the direction perpendicular to the GB ($y$), before (red circles, “simulation data”), and after (purple points) applying the correction with the electrostatic model (black line). The dashed vertical lines indicate the GB regions.

As shown in Figure 7, the “simulation” data points of the uncorrected formation energy follow the saw-tooth model function. The data points in the bulk grain regions deviate from the model on the order of $0.01$ eV, indicating the validity of our correction model also for supercells with periodic boundary conditions.

When the supercell containing the GB undergoes structural relaxation, the ions were observed to deviate by approximately $0.1$ Å in the GB normal direction from the center of mass of their respective plane. This makes it difficult to directly specify the two parameters $q$ and $\Delta$y in Equation (12). The potential slope depends on an averaged charge $q$ of the interface planes, and the offset between the potential lines depends on $q$ and on the position deviation $\Delta$y. The values of both parameters can be determined by fitting the model function to the simulation results of $\left[E_{\mathrm{tot}}(y) - E_{\mathrm{tot}}\left(y_{\mathrm{B}}^{\mathrm{ref.}}\right)\right]$ in the bulk regions. Using Equation (12), the fitting was carried out by minimizing the root mean square deviation between the model data and the simulation data for data points in the bulk regions, where the relative formation energies of oxygen vacancies show a linear behavior. The surface area ($A$) and the position of the reference point ($y_{\mathrm{B}}^{\mathrm{ref.}}$) remain the same as used for the unrelaxed configuration. The positions of the charged termination planes ($y_i$) are chosen as the positions of the outermost ions in each bulk region, and the parameter $\delta_{\mathrm{GB}}$ was determined accordingly.

The fitting yields $q = 2.62$ e and $\Delta y = 0.63$ Å, leading to a deviation of the simulation data points from the model function on the order of $0.01$ eV in the bulk regions. In Section 4.1 we discuss the meaning of the key parameter $q$, and how its deviation from the value of the unrelaxed supercell can be understood.

Applying this fitting procedure to the relaxed configuration of the cell containing the STGB $\Sigma$5 (210) (surface area $A = 34.10$ Å$^2$), a charge $q = 0.42$ e and a position deviation $\Delta y = 0.54$ Å were obtained. The value of $q$ only slightly deviates from $0.44$ e, the theoretical value of the unrelaxed configuration. Note that for this GB the slope of the electrostatic potential, i.e. the strength of the internal electric field, is approximately only $12$ % of the value for the STGB $\Sigma$5 (310). A detailed discussion is given in Section 4.3.

As noted above, we so far did not consider structural relaxation after the oxygen ion was removed from the cell to create the vacancy. The influence of such a relaxation on the energy of the supercell is exemplified using the relaxed configuration of the cell containing the GB $\Sigma$5 (310). In Figure 8, we plot the difference $E_{\mathrm{tot}} - E_{\mathrm{tot}}^{(0)}$ as a function of the $y$-direction for both, the relaxed and the unrelaxed supercell after creating the vacancy. Note that the position of a vacancy cannot be defined uniquely anymore after such a relaxation. We therefore took the original coordinate of the removed ion as the position of the oxygen vacancy.

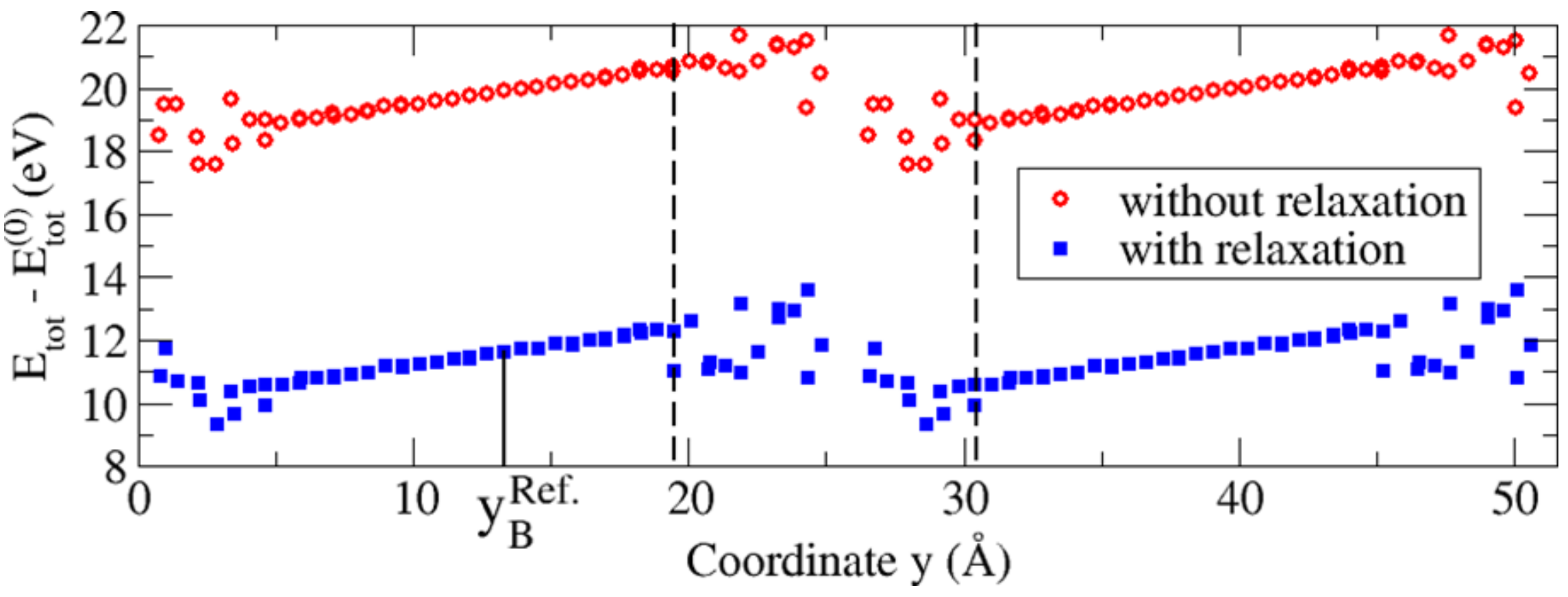

Figure 8. The energy difference $E_{\mathrm{tot}} - E_{\mathrm{tot}}^{(0)}$ without electrostatic correction in the relaxed configuration of a cell containing a STGB Σ5 (310), calculated with and without structural relaxation after forming the vacancies.

As shown in Figure 8, the data points after relaxation of the system with an oxygen vacancy (blue squares) are systematically lower (by about ~ 8.3 eV in the bulk regions) than those obtained without this relaxation (red circles). Some oxygen sites in the GB region are found to have a larger energy decrease (~ 8.6 eV), due to larger distortions of the nearest neighboring ions, which are less strongly bound at the GBs than in the bulk regions. However, relaxation does not affect the qualitative profile stemming from the electrostatic artifact. The value of the energy slope only changes on the order of $10^{-3}$ eV/Å when relaxation is done. According to Equation (12), the corresponding energy steps between the lines in the two bulk parts are then approximately equal within the same order of magnitude, too. This indicates that the strength of the interface dipole is only negligibly affected by the relaxation of the supercell containing the oxygen vacancy. This is valid for the width of the GB region, too, which we sketched by the dashed vertical lines in Figure 8. The width of the GB can be defined by the region, where the formation energy values differ from the linear behavior with respect to the $y$-coordinate. There, the local atomic environment of an oxygen site changes compared to its surroundings within the grains.

As we are primarily interested in the qualitative difference between formation energies at GBs and in bulk regions, in order to transfer such information to mesoscopic space charge models [15–18] in the continuing studies, it is sufficient to analyze data points obtained without structural relaxation of the vacancy for this purpose in this paper.

### 3.2 Corrected oxygen vacancy formation energies

The relative formation energies of positively charged oxygen vacancies at all oxygen sites after applying the electrostatic correction are plotted with respect to the $y$ coordinate in Figure 9(a) and Figure 9(b) for cells containing the STGB Σ5 (310) and the STGB Σ5 (210), respectively. Note that in both cases the ionic positions were relaxed only before the vacancy was created. In case of the GB Σ5 (310), points in the vicinity of the GB fluctuate from $-2$ eV to $1$ eV, while they fluctuate in the range $-2$ eV to $3$ eV in the vicinity of the GB Σ5 (210).

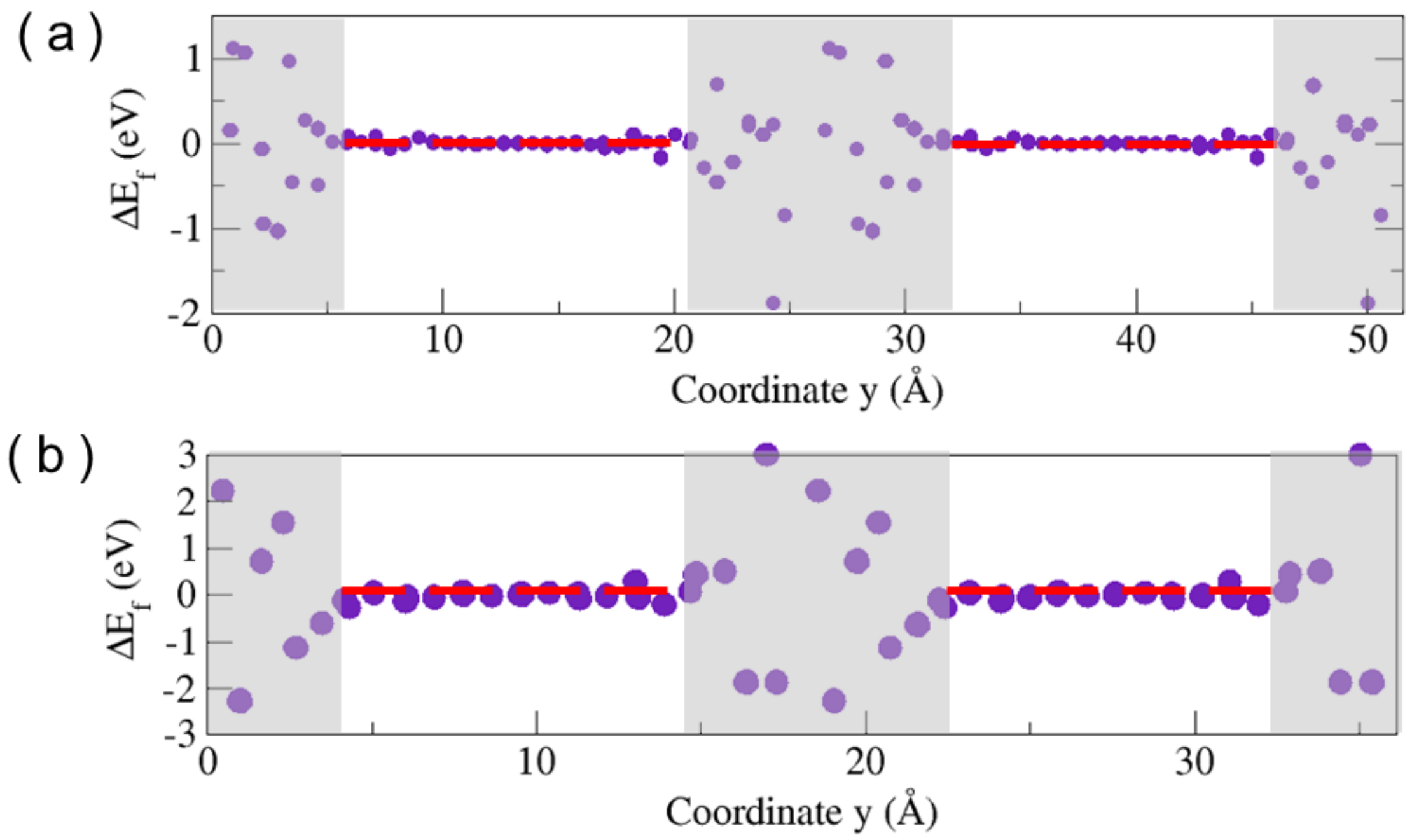

Figure 9. Corrected relative formation energies of oxygen vacancies ($\Delta E_{\mathrm{f}}$) with respect to the coordinate $y$ in the relaxed configuration of cells containing (a) the STGB Σ5 (310) and (b) the STGB Σ5 (210). The GB regions are shaded in grey. Dashed red lines mark the averaged relative formation energies in the bulk regions of the grains, which are approximately zero by construction.

# 4 Discussion

## 4.1 Comparison between unrelaxed and relaxed configurations

The surface charge density of the termination planes in the relaxed configurations, which can be obtained from fitting the model functions of the internal electrostatic potential to the simulation data (*cf.* Section 3.1.2), can be deduced by comparing the repeated units of relaxed and unrelaxed bulk configurations. This is explained for the example of the cell containing a STGB Σ5 (310). As described in Section 2.1, the corresponding unrelaxed configuration consists of repeated pairs of the negatively charged O-O plane ($-q$) and the positively charged Sr-Ti-O plane ($q$), with surface charges of $-2.80$ e and $2.80$ e, respectively. The planes in the bulk regions are equally spaced by a distance $d = 0.618$ Å. Such a repeated unit is sketched in Figure 10(a) (dashed vertical lines). The corresponding electrostatic potential is indicated by red solid lines [*cf.* Figure 3(b)].

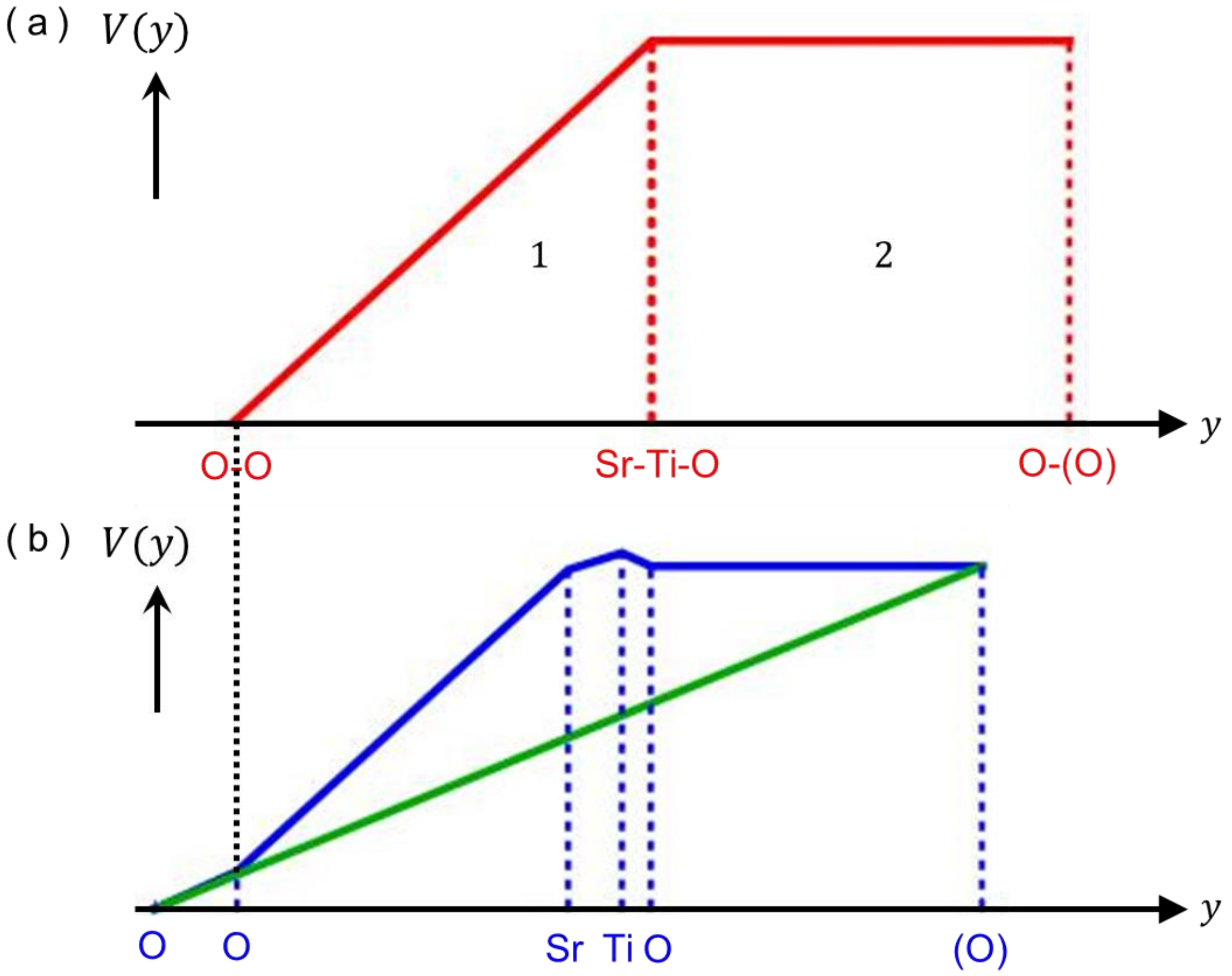

Figure 10. The electrostatic potentials of a repeated unit within a cell containing a STGB Σ5 (310) for (a) the unrelaxed configuration and (b) the relaxed configuration. Positions of the O-O and Sr-Ti-O planes in the unrelaxed configuration are indicated by dashed red vertical lines. They lead to the electrostatic potential sketched by the solid red line. The positions of the ions in the relaxed configuration deviate from these planes, as shown by dashed blue vertical lines. The deviations are exaggerated here for clarity. The second O ion in the sequence is located at the same position as the O-O plane in the repeated unit of the unrelaxed configuration, as shown by the dashed black vertical

line. The corresponding potential is given by the solid blue line. Its macroscopic average is indicated by the solid green line.

However, the relaxed ionic positions in the repeated unit deviate from the corresponding unrelaxed ionic positions, as marked by the dashed blue vertical lines in Figure 10(b). Ions within the repeated unit of the relaxed configuration are observed to preserve the same sequence everywhere in the bulk regions: O-O-Sr-Ti-O-(O). The second O ion in this sequence is located at the same position as the O-O plane in the repeated unit of the unrelaxed configuration (the correspondence is shown with the black dashed vertical line in Figure 10). This position is set as the reference point. The third O ion in the sequence is then located almost at the same position as the Sr-Ti-O plane in the repeated unit of the unrelaxed configuration. The last O in the sequence (in parenthesis) represents again the first O-ion in the sequence of the following unit. Using the averaged ionic distances, the ions are located in the GB normal direction ($y$-direction) with respect to the reference point at: $-0.012$ Å, $0$ Å, $0.575$ Å, $0.597$ Å, $0.614$ Å and $1.266$ Å. The induced electrostatic profile (solid blue line) can be calculated by summing up the electrostatic potentials from each ion (lattice plane) according to Equation (8). Using a dipole to macroscopically describe this electrostatic profile [green line in Figure 10(b)], the resulting effective dipole charge is calculated as $2.62$ e from Equation (9), which is exactly the value of the surface charge $q$ obtained from the fitting of the electrostatic model to the simulation data in Section 3.1.2 (Figure 7).

The same conclusion holds for the repeated units within the cells containing the STGB $\Sigma 5$ (210), where the sequence of ions in the bulk regions is O-Ti-O-Sr-O-(O). Using the averaged ionic distances and referencing again to the second ion in the sequence (Ti ion), they are located in $y$-direction at: $-0.121$ Å, $0$ Å, $0.059$ Å, $0.817$ Å, $0.867$ Å and $1.647$ Å. Averaging the electrostatic profile introduced by this sequence yields the effective dipole charge of $0.42$ e in perfect agreement with the fitting result (*cf.* Section 3.1.2).

## 4.2 Analysis of the oxygen vacancy formation energies

The formation energies of oxygen vacancies in the cells containing the two considered STGBs show similarities, as displayed in Figure 9: oxygen sites with positive and negative $E_\mathrm{f}$ values relative to the values in the bulk region of the grains, accumulate in the vicinity of the grain boundaries. Some sites are beneficial by around $-1$ to $-2$ eV, indicating that oxygen vacancies prefer these specific sites near the charged interface planes.

Oxygen vacancy formation energies are further analyzed with respect to the local atomic environments of their respective sites. A descriptor of the local environment should include factors like the coordination number, neighbor species, and atomic distances in a defined range, and thereby it should reflect the site energy. To quantify these three factors in a compact form, we define a local charge density $\rho$ as follows:

$$\rho = \sum_i^N \frac{q_i}{(4/3)\pi r_i^3}. \tag{15}$$

where $N$ is the number of neighbor atoms in the local environment. $q_i$ is the charge of the neighboring atom $i$ and $r_i$ is its distance to the selected oxygen site.

The range of the local environment was first evaluated applying a cut-off radius $r_0 = 2.76$ Å, which is the interionic distance of an oxygen site to its nearest-neighbor oxygen site in the perfect STO structure. However, neighboring ions slightly beyond this cut-off radius were observed to influence the charge density considerably. For example, for the oxygen sites at the STGB Σ5 (310) in the relaxed configuration, one neighboring ion at a distance of $\sim 0.1$ Å above the cut-off radius changes the charge-density value by around 10 %. To obtain more reliable values, we extended the cut-off radius by a factor of 1.3, and additionally softened it by applying a linear interpolation function $f(r)$ for a fractional counting [43] in the extended range:

$$f(r) = \frac{1.3r_0 - r}{1.3r_0 - r_0},\ r_0 \le r \le 1.3r_0. \qquad (16)$$

From Equations (15) and (16), the local charge density was calculated for each oxygen site, and used to analyze the relative formation energies with respect to their local environments, as displayed in Figure 11.

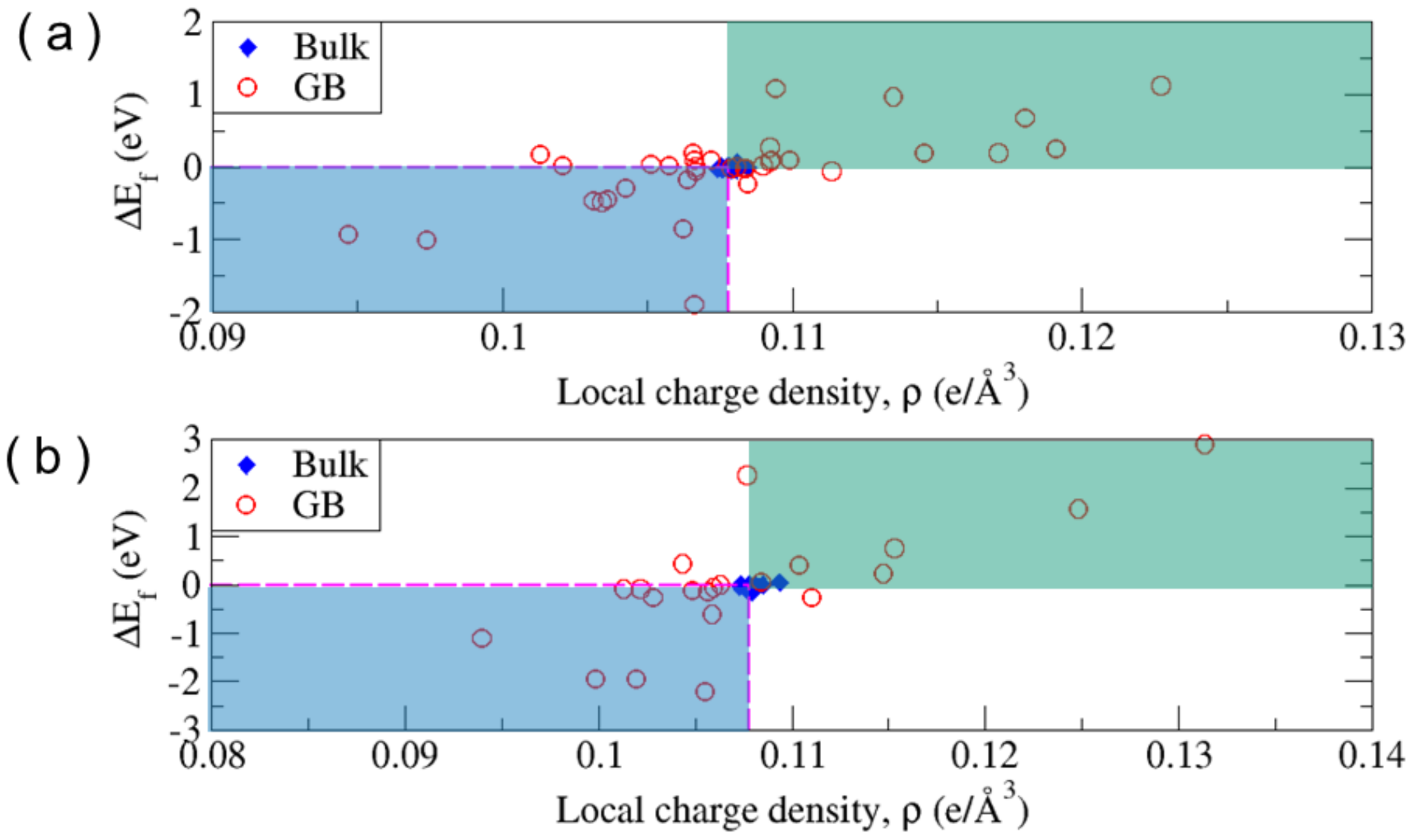

Figure 11. Relative formation energy of oxygen vacancies ($\Delta E_f$) plotted with respect to the local charge density for the relaxed configurations of cells containing (a) a STGB Σ5 (310), and (b) a STGB Σ5 (210). The intersection of the dashed pink lines labels the reference point calculated for an oxygen site within the bulk of a perfect, i.e. defect-free STO crystal. Data points shown by blue diamonds represent bulk-like oxygen sites while red circles mark the oxygen sites near the GBs.

Here, oxygen sites in bulk and GB regions are distinguished by blue diamonds and red circles, respectively. A reference data point, calculated for an oxygen site in a perfect STO crystal with charge density $\rho_0 = 0.108\ e/\text{Å}^3$, is labeled by the intersection of the dashed pink lines. Data points of bulk-like sites are located close to this reference point. Data points for GB-like sites can be distributed in two regions: the blue-shaded region contains data of oxygen vacancies with negative relative formation energy and a smaller charge density at the respective site ($\rho < \rho_0$) than in the bulk. The green-shaded region contains data of oxygen vacancies with positive relative formation energy and larger charge density at the respective site ($\rho > \rho_0$) than in the bulk. These results indicate that oxygen vacancies tend to be trapped at specific sites in the vicinity of the GBs, where the local charge densities have lower values than in the bulk.

### 4.3 The electrostatic artifacts in low-angle $(\mathbf{i10})$ tilt grain boundaries

So far, we studied the electrostatic artifacts in cells containing the two STGBs Σ5 (310) and Σ5 (210), which are characterized by the tilt angles $\theta = 18.4°$ and $26.6°$, respectively. However, internal electric fields are not only present in cells containing these two high-angle tilt grain boundaries, but are expected to appear in low-angle tilt grain boundaries, as well. To illustrate this, we investigate low-angle tilt grain boundaries with orientations $(i10)$ for $i = 6, 7, 8, 9, 12, 13, 20, 21, 40, 41$.

According to Equations (10) and (12), the slope of the electrostatic potential in the bulk regions of supercells containing STGBs, and hence the strength of the internal electric field, is proportional to $\frac{q}{2A}$. $q$ denotes the charge of a lattice plane, which can be obtained by summing up the charges of all ions on this plane (*cf.* Figure 1). The composition of planes in $SrTiO_3$ has been discussed in the Appendix A of Ref. [34]: Planes of the form $(i10)$, where $i$ is an odd integer, are composed of Sr-Ti-O or O-O. If $i$ is an even integer, the planes have the composition Sr-O or Ti-O-O. Note that in the following discussions, the values of $q$ refer to their absolute which are listed in Table 2. We analyzed $q$ values for both, the effective charges of the species used in the Thomas potential and the formal charges.

Table 2. Formal and Thomas charges of lattice planes for $(i10)$ oriented GBs for odd and even integer values of $i$

| $q$ [e] | $i$ odd | $i$ even |
|---|---|---|
| Thomas | 2.80 | 0.44 |
| Formal | 4.00 | 0 |

The interface area $A$ of one repeated unit can be derived as $a_{\mathrm{STO}}^2\sqrt{i^2+1}$. With the substitution $\tan\theta = 1/i$, we obtain $A = a_{\mathrm{STO}}^2/\sin\theta$. For low angles $\theta$, the sine function can be approximated: $\sin\theta \approx \theta$. It follows for the proportionality factor of the strength of the internal electric field:

$$\frac{q}{2A} \approx \frac{q}{2a_{\mathrm{STO}}^2}\theta. \tag{17}$$

Hence, the strength of the internal electric field is proportional to the tilt angle in the low-angle $(i10)$ tilt grain boundaries. This is further confirmed by plotting $\frac{q}{2A}$ with respect to the tilt angle $\theta$ as shown in Figure 12. Five orientations with oddly numbered $i$ $(7, 9, 13, 21, 41)$ and five with evenly numbered $i$ $(6, 8, 12, 20, 40)$ were randomly selected. They are represented by solid and hollow symbols, respectively. Additionally, for each orientation, the value of $\frac{q}{2A}$ evaluated from Thomas charges and formal charges are distinguished, leading altogether to four lines corresponding to the four plane charges listed in Table 2.

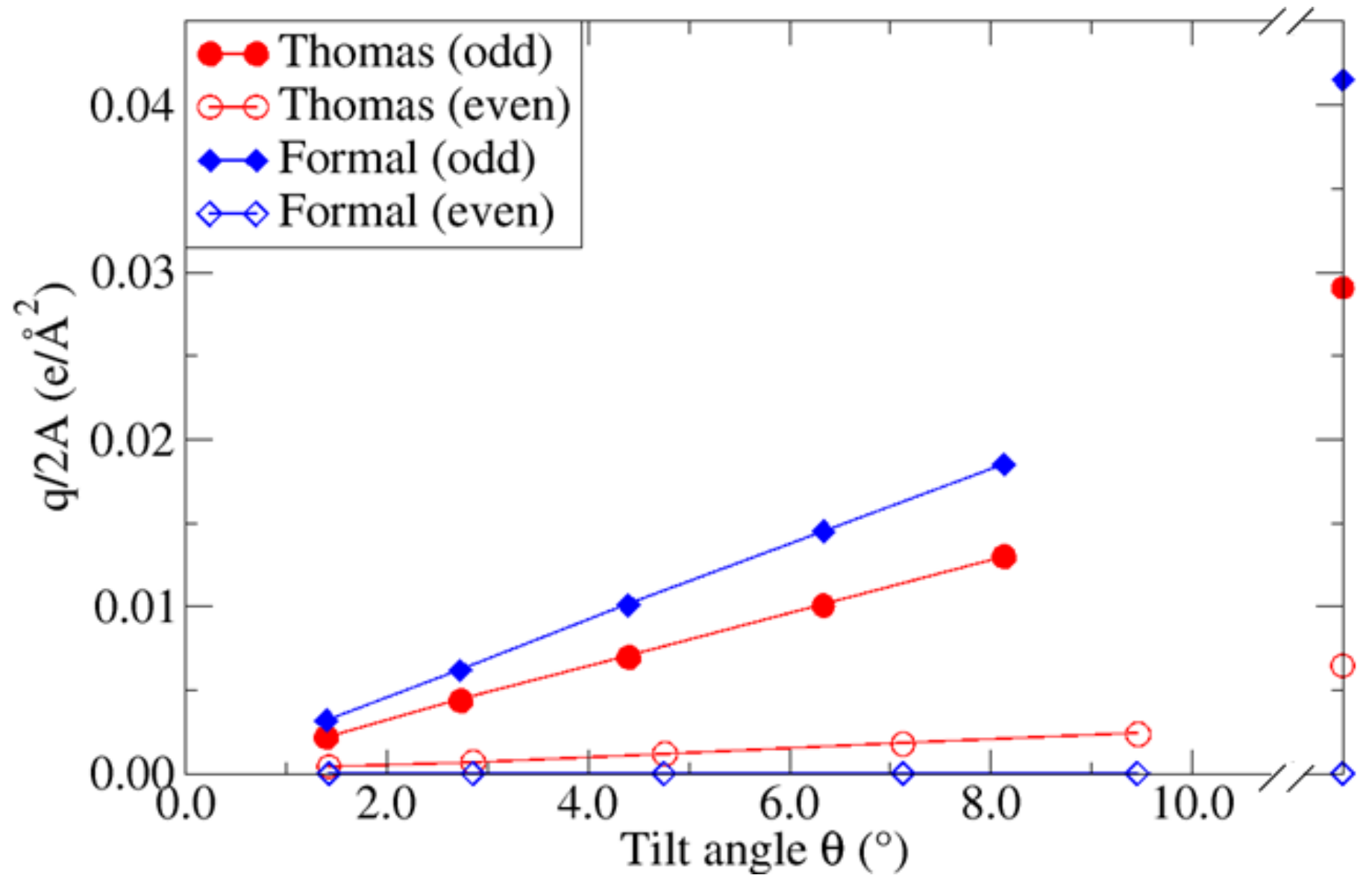

Figure 12. The values of $\frac{q}{2A}$ for $(i10)$ oriented low-angle grain boundaries (with odd $i = 7, 9, 13, 21, 41$ and even $i = 6, 8, 12, 20, 40$) plotted with respect to the tilt angle. Data points represented by red circles are calculated with the Thomas charges, and those represented by blue diamonds with formal charges. The lower two data points and the higher two data points marked at the right $y$-axis correspond to the $(210)$ and to the $(310)$ oriented grain boundaries, respectively, which we studied in this work.

For comparison, data points for systems containing $(210)$ and $(310)$ oriented grain boundaries are included accordingly at the right $y$-axis. The calculated values of $\frac{q}{2A}$ are relatively small in the small angle region, which means that the corresponding artificial fields are likely small enough to be neglected in atomistic GB simulations. This analysis can be transferred to other tilt grain boundaries, e.g. those with indices $(i20)$, $(i30)$ and so on, to which different plane charges can be assigned. However, the linear approximation given in Equation (17) always holds for a small $\theta$.

As can be seen in Figure 12, the existence of an internal field is sensitive to the set of charges used for rigid ions in atomistic GB simulations. When formal charges are applied to the cases of evenly numbered $i$ for the $(i10)$ oriented GBs, the planes are charge neutral and therefore no electrostatic potential appears in the supercell [34].

No internal electrostatic fields were reported by Ramadan and De Souza [25], who calculated oxygen vacancy formation energies at 13 different low-angle $[100]$ $(01i)$ tilt grain boundaries with odd and even $i$ in STO. They also applied a rigid-ion model for STO, adopting ionic charges which were originally derived for binary oxides by Pedone *et al.* [44], namely $+1.2\mathrm{e}$, $+2.4\mathrm{e}$ and $-1.2\mathrm{e}$ for Sr, Ti, and O, respectively. These values are different from the formal charges, but also add up to zero on lattice planes corresponding to grain boundaries with even $i$, which makes it obvious from Figure 12, that no internal fields were detected in these cases. However, on grain boundary planes with odd $i$, the charges used by Ramadan and De Souza [25] add up to $2.4e$, in which case an electric field must have been present in the cells. But firstly, low-angle grain boundaries were studied there, with (following our definition) a maximum tilt angle of 11.3° ($i = 5$). Secondly, those were set up in cells with lengths perpendicular to the grain boundaries of about three times as long as the length of the STGB Σ5 (310) cell we considered in our work. Since this cell dimension enters the formula for the magnitude of the potential slope [Equation (12)] in the denominator, both aspects together lead to internal field strengths of about one order of magnitude lower than the field values we obtained for the STGB Σ5 (310). If they detected it at all by their analysis, such small fields might have been considered negligible by Ramadan and De Souza [25].

Oyama *et al.* [8] studied oxygen vacancy formation energies across STO supercells for both, the STGBs Σ5 (310) and Σ5 (210), applying the Thomas potential with formal charges. They did not mention internal electric fields in their cells. While in the case of STGB Σ5 (210), this is in agreement with our analysis (*cf.* Figure 12), there should have been a noticeably large field in the supercells of Oyama *et al.* [8] containing the STGB Σ5 (310), even though the cell size perpendicular to the grain boundary was about twice as long as the one we considered in our work.

# 5 Summary and Conclusions

In this work, we investigated formation energies of oxygen vacancies in supercells containing symmetric tilt grain boundaries of the form Σ5 (310) and Σ5 (210) in $SrTiO_3$ by performing atomistic molecular-statics simulations with a rigid-ion potential. The following conclusions can be drawn:

1. In the ionic model, an internal electric field is present inside the supercells containing a GB with broken mirror symmetry, when the lattice planes are charged. This effect is considered as an electrostatic artifact, and should be corrected in energy calculations such as the determination of the formation energy of an oxygen vacancy in our study. Especially for GBs with large tilt angles, strong fields can be present within the simulation cells.

2. A continuum model was proposed for the defect calculations to correct the electrostatic effects in supercells for both, open and periodic boundary conditions. This model requires only two free parameters, namely the surface charge $q$ and a position deviation $\Delta y$. These values can directly be given in the case of ideal, i.e. symmetrically set up and unrelaxed supercells. Structural relaxation leads to an internal polarization, in which case $q$ and $\Delta y$ can be obtained by a fitting of the derived model functions to the simulation data.

3. The corrected formation energies indicate the effect of GBs to attract oxygen vacancies in polycrystalline $SrTiO_3$. We found a relationship between the local charge density around an oxygen site and its formation energy. Oxygen vacancies are trapped at specific sites in the vicinity of the GBs, where the local charge densities have lower values than in the bulk.

The present work did not take any charge compensation effects for the charged GBs, e.g. by various types of point defects in space-charge zones, into consideration. This requires continuing studies to transfer information obtained in this atomic-scale study to mesoscopic space charge models. The developed correction method has the potential to be extended to more complicated interfaces such as asymmetric tilt grain boundaries (ATGBs). Experimentally, ATGBs are more frequently observed than STGBs in ceramic $SrTiO_3$ microstructures [3,45]. However, atomistic structures of ATGBs have only been occasionally studied [46]. The correction method presented here offers the potential of extending atomistic simulations of charged point defects to more general interfaces in ceramics. Even though developed for a rigid-ion potential in this work, the method can be applied to systems treated by DFT as well. It extends the model developed by means of DFT calculations of charged surfaces in slabs [21] to supercells with interfaces and interface dipoles. The existence and consequences of interface dipoles have been extensively studied by DFT methods in the past, going back to the work of Louie and Cohen [47] followed by many publications ever since [48,49].

**Acknowledgements**

This work was funded by the German Research Foundation (DFG); Grants No. MR22/6-1 and EL155/31-1 within the priority programme “Fields Matter” (SPP 1959). The authors thank Prof. Roger De Souza (RWTH Aachen) for the helpful discussions. Computations were carried out on the bwUniCluster computer system of the Steinbuch Centre of Computing (SCC) of the Karlsruhe Institute of Technology (KIT), funded by the Ministry of Science, Research, and Arts Baden Württemberg, Germany, and by the DFG. Structure figures were prepared with VESTA [50].